\documentclass[fleqn,usenatbib]{mnras}
\usepackage{newtxtext,newtxmath}
\usepackage[T1]{fontenc}
\DeclareRobustCommand{\VAN}[3]{#2}
\let\VANthebibliography\thebibliography
\def\thebibliography{\DeclareRobustCommand{\VAN}[3]{##3}\VANthebibliography}
\usepackage{graphicx}	
\title[First ALMA Observation of a Solar Prominence]{First High Resolution Interferometric Observation of a Solar Prominence With ALMA}
\author[N. Labrosse et al.]{
Nicolas Labrosse,$^{1}$\thanks{E-mail: Nicolas.Labrosse@glasgow.ac.uk}
Andrew S. Rodger,$^{1}$
Krzysztof Radziszewski,$^{2}$
Pawe\l{} Rudawy,$^{2}$
Patrick Antolin,$^{3}$
\newauthor
Lyndsay Fletcher,$^{1,4}$
Peter J. Levens,$^{1}$
Aaron W. Peat,$^{1}$
Brigitte Schmieder$^{1,5}$
and 
Paulo J. A. Sim\~oes$^{6,1}$
\\
$^{1}$SUPA, School of Physics \& Astronomy, University of Glasgow, Glasgow G12 8QQ, UK\\
$^{2}$ Astronomical Institute, University of Wroc\l{}aw, PL\\
$^{3}$Department of Mathematics, Physics and Electrical Engineering, Northumbria University, Newcastle Upon Tyne, NE1 8ST, UK\\
$^4$Rosseland Centre for Solar Physics, University of Oslo, P.O.Box 1029 Blindern, 0315 Oslo, NO\\
$^{5}$Centre for mathematical Plasma-Astrophysics, KU Leuven, Leuven, BE\\
$^{6}$Centro de Rádio Astronomia e Astrofísica Mackenzie, Escola de Engenharia, Universidade Presbiteriana Mackenzie, São Paulo, BR
}

\date{Accepted XXX. Received YYY; in original form ZZZ}

\pubyear{2022}

\begin{document}
\label{firstpage}
\pagerange{\pageref{firstpage}--\pageref{lastpage}}
\maketitle

\begin{abstract}
We present the first observation of a solar prominence at $84-116$~GHz using the high resolution interferometric imaging of ALMA. 
Simultaneous observations in H$\alpha$ from Bia\l{}kaw Observatory and with SDO/AIA reveal similar prominence morphology to the ALMA observation. The contribution functions of 3~mm and H$\alpha$ emission are shown to have significant overlap across a range of gas pressures. We  estimate the {maximum} millimetre-continuum optical thickness to be $\tau_\mathrm{3mm}\approx 2$, {and the brightness temperature from the observed H$\alpha$ intensity}. The brightness temperature {measured by ALMA} is $\sim 6000-7000$~K in the prominence spine, {which correlates well with the estimated brightness temperature for a gas temperature of 8000~K}. 
\end{abstract}




\section{Introduction}

Solar prominences are one of the most striking illustrations of how the magnetic field shapes the atmosphere of our Sun \citep{review,iauproc,2015ASSL..415.....V}, and constitute the best  example in the solar system of natural plasma confinement: the magnetic field somehow supports a cool ($T\sim 10^4$~K), dense ($n_e \sim 10^{10}$~cm$^{-3}$) plasma in a hot ($T\sim 10^6$~K), tenuous ($n_e \sim 10^8$~cm$^{-3}$) environment. 
If the magnetic field loses its equilibrium configuration, the prominence may become unstable and erupt,  throwing vast amounts of magnetized plasma into the heliosphere. These prominence eruptions are often associated with the most energetic events in the solar system: flares and coronal mass ejections.
Understanding their formation, and their temporal evolution, is relevant for our understanding of the interplay between plasma and magnetic field in stellar coronae.
Attempts at inferring the temperature structure of solar prominences have been limited due to the techniques used and available spectral windows \citep{2014LRSP...11....1P}. Yet, the gas temperature is a fundamental quantity to measure to understand the  lifecycle of solar prominences, from their formation, to their evolution and their disappearance \citep{2016SSRv..200....1W}.

The Atacama Large Millimeter/submillimeter Array (ALMA) offers a new window on the solar prominence plasma. The formation mechanism of the sub-mm radiation means that brightness temperature measurements are in principle a good proxy for the electron temperature \citep{1985ARA&A..23..169D}. We carried out the first observations of a solar prominence in April 2018 (ALMA Cycle 5).
Section~\ref{sec:prom_obs} describes the observations and data reduction procedures for  ALMA, ground-based H$\alpha$, and SDO/AIA. 
Section~\ref{sec:prom_results} presents a millimetre-continuum optical thickness diagnostic using the H$\alpha$ integrated intensity, the  brightness temperature results,  and a discussion of the prominence morphology as seen by ALMA compared to H$\alpha$ and AIA.
Finally, Section~\ref{sec:prom_conc} presents our conclusions.

\section{Observations and Data Reduction}\label{sec:prom_obs}

The small prominence appearing above the south-{west} solar limb on 19th April 2018 was well observed by ALMA together with other ground-based and space-based observatories. \citet{2021A&A...653A...5P} carried out a detailed analysis of its emission in the Mg~II h and k lines observed by the IRIS spacecraft, while \citet{2021A&A...653A..94B} investigated its dynamics in relation to its magnetic structure. 

ALMA observed off the solar limb near the heliocentric coordinates ($+650\arcsec,-750\arcsec$) between 15:20 and 17:45~UTC on the $\mathrm{19^{th}}$ of April $2018$.
The mode of operation planned for this observation was a mosaic with a proposed total field of view (FOV) of $40$\arcsec\ by $80$\arcsec.
This FOV was produced by cycling through a total of $5$ pointings, each with a smaller FOV determined by the observing wavelength and array configuration. 
In this case the observing band was Band~$3$ ($84$--$116$~GHz) and the array configuration was C$43$-$3$, the largest  array configuration available for solar observations at the time, giving the highest possible spatial resolution.
This prominence was also observed two days earlier in ALMA Band~6. Here we focus our analysis on ALMA Band~3 data as they offer a better diagnostic potential for solar prominences \citep{2017SoPh..292..130R}.

\subsection{ALMA Band~$3$}\label{sec:ALMA_prom_obs}

The ALMA observation was split into two main observing blocks: from 15:38 -- 16:32 (Block~$1$) and from 16:50 -- 17:44 (Block~$2$). 
Within each block ALMA conducted $7$ scans of the total solar FOV, with off-target calibration occurring between each scan. 
The average time taken for each scan was $\sim6$~minutes.

Issues with the system temperature were found in several of the 43 12m array antennas where their XX and YY {polarizations} were too large, resulting in very large estimates for the noise in the synthesised images.
To combat this, the problem antennas were identified, which included $8$ antennas for Block~$1$ and $10$ for Block~$2$, which were subsequently removed from the image synthesis. 
Whilst the reliability of the data is improved by removing the antennas, the resolution suffers somewhat due to the reduction in overall baseline number. 

\begin{figure}
	\centering
	\includegraphics[width=\columnwidth]{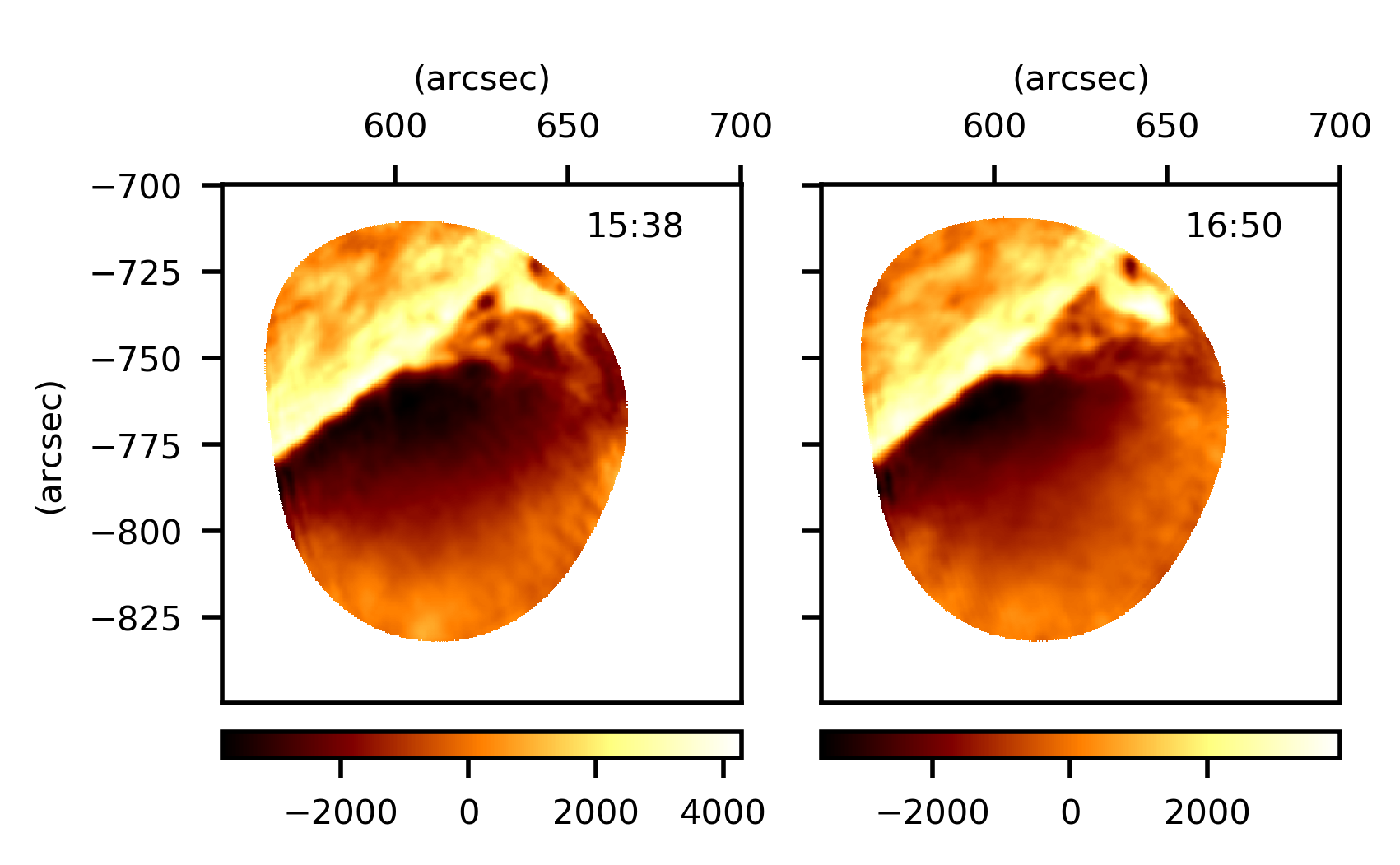}
	\caption{Synthesised ALMA interferometric images for Blocks 1 and 2 of the prominence on 19 April $2018$. Each image is produced by averaging over 7 scans ($\sim 42$~min), with the start time of each block given in the image titles. 
	The colourbars show the relative brightness temperature in Kelvin.}
	\label{fig:interferometric_prom}
\end{figure}

To produce an image from the imperfectly sampled visibility data, we used the deconvolution algorithm of the \emph{Common Astronomical Software Applications} (CASA) program \textit{tclean} with the multiscale extension to the CLEAN algorithm \citep{Hogbom} presented in \citet{multiscale}. 
The resultant cleaned interferometric images for the whole time range are shown in Figure~\ref{fig:interferometric_prom}.

Interferometric images do not yield absolute brightness temperatures, but rather relative brightness temperature values around some background value determined by the largest spatial scales. 
To produce absolute brightness temperature images the interferometric images must be combined with single dish images to provide measurements of the total flux in the image \citep{Koda2011}. 
The process  used here is to combine the interferometric and the single dish measurements from the Total Power (TP) array -- this is called ``feathering''.
Feathering involves combining two images together through the weighted sum of the Fourier transform components in the \emph{uv}-plane.
This will also provide the low spatial resolution component of the image due to the limitations in the shortest possible baseline in the interferometer array. 
Across the entire observation ALMA conducted a total of $14$ observing scans of the entire solar disc, each taking $\sim10$~minutes. 
The method for TP full-disc observations is described in \citet{White2017}. 
For each interferometric synthesised image a corresponding TP image was chosen to feather with it to produce an absolute brightness temperature image. 
The choice of corresponding TP image was by either having the longest period of co-temporal observation, or if that was not possible, the shortest difference in time between it and the synthesised interferometric image.

\subsection{H$\alpha$}

The prominence was observed in the H$\alpha$ line using the Large Coronagraph and the Multi-Chanel Subtractive Double Pass Spectrograph (MSDP) located at Bia\l{}kow Observatory (University of Wroc\l{}aw in Poland). The MSDP spectrograph measured the intensity at $23$ different wavelengths ranging from $-1.1$~\AA\ to $+1.1$~\AA\ from the line centre. 
The MSDP quasi-monochromatic images used in the analysis were obtained from a combination of 13 sub-images during observing in slow-mode using a scanner. A total of $739$ scans were taken between 10:02 and 15:54~UTC. 
In this study we use a singular scan from this observation to estimate the optical thickness of the Band~$3$ emission {which is largely dominated by the free-free thermal continuum}.
This scan was recorded between 15:35:08 and 15:35:24~UTC, roughly $3$ minutes before the start of the first ALMA observation. 
The data has been calibrated in absolute units and the scattered light was subtracted \citep{2007A&A...461..303R}.

\subsection{AIA}

In Section~\ref{sec:AIA_coaligned} we compare the image of the prominence synthesised with ALMA in Band~$3$ across observing Block~$2$ to simultaneously observed imaging from the \emph{Atmospheric Imaging {Assembly}} (AIA) on board the \emph{Solar Dynamics Observatory} (SDO) \citep{Lemen2012}.
As the ALMA image for Block~$2$ is time-averaged,  small scale motions will be undetectable. 
To allow for direct comparison with this ALMA image and the prepped AIA images, the AIA images are also averaged over the corresponding time range. 
In AIA bands where the prominence signal is faint, or there was a low {signal-to-noise ratio}, this time averaging produced a more distinct view of the prominence to compare to the ALMA image. {AIA data where the signal was strong, e.g. at $304$~\AA, were also averaged with respect to time so that any time-dependent fine structure will be unresolved in a similar manner to the lower temporal resolution ALMA image.}

\section{Results and Analysis}\label{sec:prom_results}

\subsection{H$\alpha$  Intensity and the Millimetre-Continuum Optical Thickness}

\begin{figure}
	\centering
	\includegraphics[width=0.9\columnwidth]{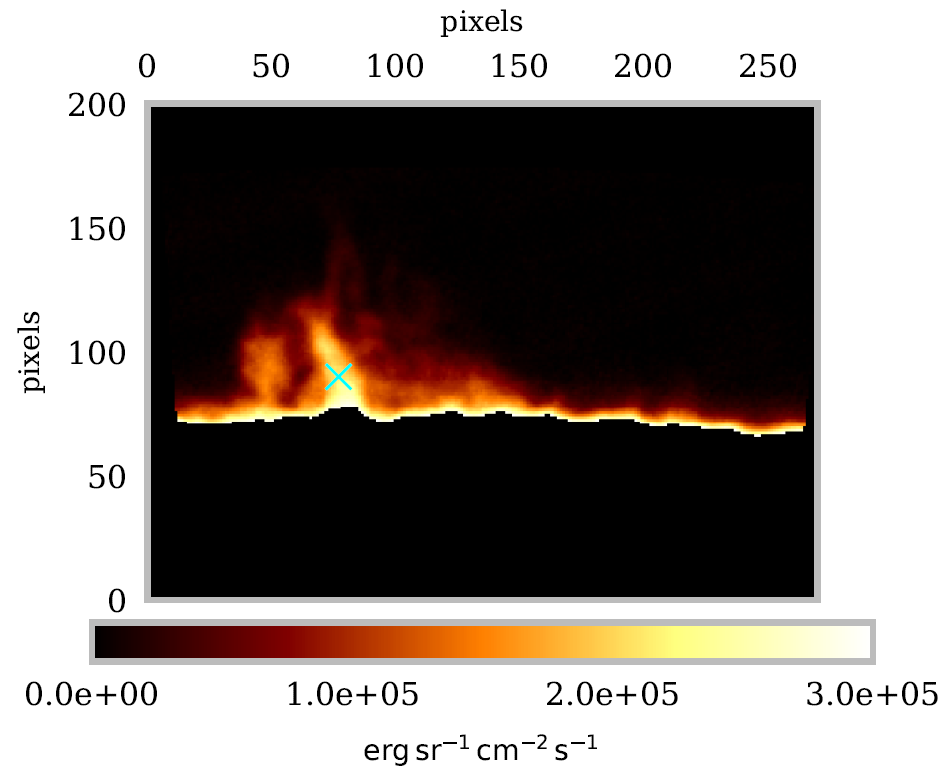}
	\caption{Prominence observed in H$\alpha$ integrated intensity by the MSDP spectrograph.
	The cyan cross marks the centre of the prominence, where $E({\mathrm{H_\alpha}})\approx 2.2\times 10^5$~erg s$^{-1}$ cm$^{-2}$ sr$^{-1}$.
}
	\label{fig:Halpha_integrated_intensity}
\end{figure}

The first step to estimate the optical thickness of the prominence in ALMA Band~$3$  is to calculate the integrated intensity across the H$\alpha$ line profile. The result is shown in Figure~\ref{fig:Halpha_integrated_intensity}. 

\begin{figure}
	\centering
	\includegraphics[width=\columnwidth]{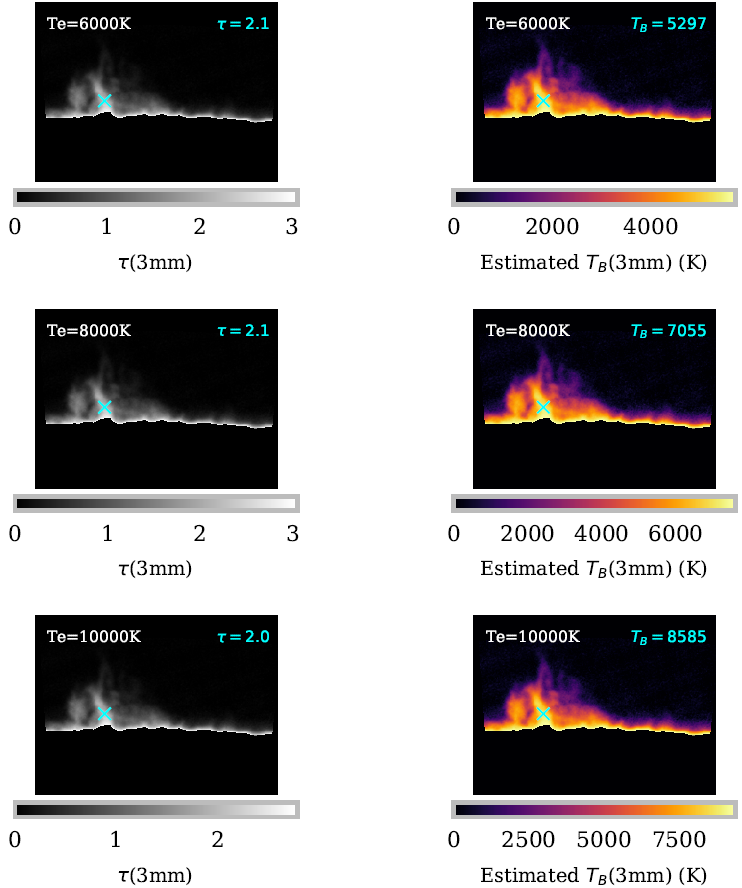}
	\caption{Optical thickness and brightness temperature estimations for ALMA Band~$3$ emission as calculated using the H$\alpha$ integrated intensities from Figure~\ref{fig:Halpha_integrated_intensity} and an isothermal assumption. 
		The estimation was calculated at $3$ constant temperatures, where the values for the $f$-factor (Equation~\ref{eq:f_factor}) were the same as quoted in \citet{2015SoPh..290.1981H}.}
	\label{fig:prom_tau_est}
\end{figure}

The optical thickness of the mm-continuum is related to the H$\alpha$ integrated intensity $E({\mathrm{H_\alpha}})$ for an isothermal line of sight (LOS) through 
\begin{equation}
	\tau_{\nu} \approx 4.55\times 10^{17} g_{\mathrm{ff}} e^{-T_0/T}E({\mathrm{H_\alpha}})/(\nu^2b_3(T)) \propto g_{\mathrm{ff}}(\nu,T)f(T) \ ,
\end{equation}
where $g_{\mathrm{ff}}$ is the {Gaunt} factor, $T$ the temperature, {$T_0=17534$~K}, $\nu$ the frequency, {$b_3(T)$ the LTE (Local Thermodynamic Equilibrium) departure coefficient of the third energy state of hydrogen \citep[varying from 0.75 at 6000~K to 2.97 at 10000~K according to][]{Jejcic2009},} 
and $f(T)$ is defined by \citep{2015SoPh..290.1981H,2019PhDT.......103R}
\begin{equation}\label{eq:f_factor}
	f(T) =  {e^{-T_0/T}}/{b_3(T)}.
\end{equation}
In \citet{2015SoPh..290.1981H} the authors use $3$ values for the $f(T)$ factor calculated using $1$D prominence models at constant temperatures of $6000$, $8000$ and $10000$~K.
In fact, the variation with temperature of the optical thickness estimation using H$\alpha$ integrated intensity is fairly small for temperatures up to $\sim 20000$~K. 
We use the same values for $f(T)$ as quoted in \citet{2015SoPh..290.1981H} and the resulting optical thickness and brightness temperature estimations are shown for each of $6000$, $8000$ and $10000$~K constant temperature plasmas in Figure~\ref{fig:prom_tau_est}.

From Figure~\ref{fig:prom_tau_est}, it can be seen that if the observed prominence were to have a constant temperature between $6000$ and $10000$~K, the optical thickness of the mm-continuum should lie {in the densest parts} between $\approx1$ and $3$, with the value at the centre of the prominence spine (cyan cross) at $\approx2$, {and going below 1 at the edges}. 
An optical thickness of $2$ in Band~3 has been shown to be too low to make a direct electron temperature diagnostic from the mm-continuum brightness temperature for isothermal prominence models \citep{2017SoPh..292..130R}.

If the temperature of the plasma is higher than the $10000$~K considered here, which is perhaps unlikely due to the observed H$\alpha$ intensities, the optical thickness of the mm-continuum could be higher due to an increased magnitude of the $f$ factor.
Our calculations (see next section) show that, even for the unlikely case where the prominence plasma is up to $20000$~K in temperature, the optical thickness of the mm-continuum would not exceed $\approx4-5$. 

\subsection{Time Averaged ALMA Images}

Figure~\ref{fig:prom_tau_est} suggests that the brightness temperature appears to be  uniformly distributed across the width of the prominence spine. 
The prominence models of \citet{2017SoPh..292..130R} indicate that the mm-continuum emission  peaks in lines of sight passing through optically thick material from the prominence-to-corona transition region (PCTR). This suggests  that the PCTR is either optically thin in this observation or that the PCTR is not resolved, or both.
The resolution of the ALMA maps is defined by the semi-major and semi-minor axes of the synthesised beam, which is $1.98$\arcsec\ x $1.52$\arcsec\ for Block~$1$ and $2.17$\arcsec\ x $1.56$\arcsec\ for Block~$2$. 
The visibility of fine structures is also affected by the long time range used in the production of this image as well as the frequency bandwidth, as emission at different frequencies will be produced from different layers in the fine structures.

Our estimated maximum optical thickness of the prominence spine at 3~mm ($\sim2$) is consistent with the hypothesis that LOSs going through the sparser PCTR material are optically thin in this observation. 
{As an optical thickness of $2$ is below} that which is required for a direct measurement of the local electron temperature of the emitting plasma, an absolute brightness temperature measurement  presents a lower boundary to the electron temperature measurement. 

\begin{figure}
	\centering
	\includegraphics[width=\columnwidth]{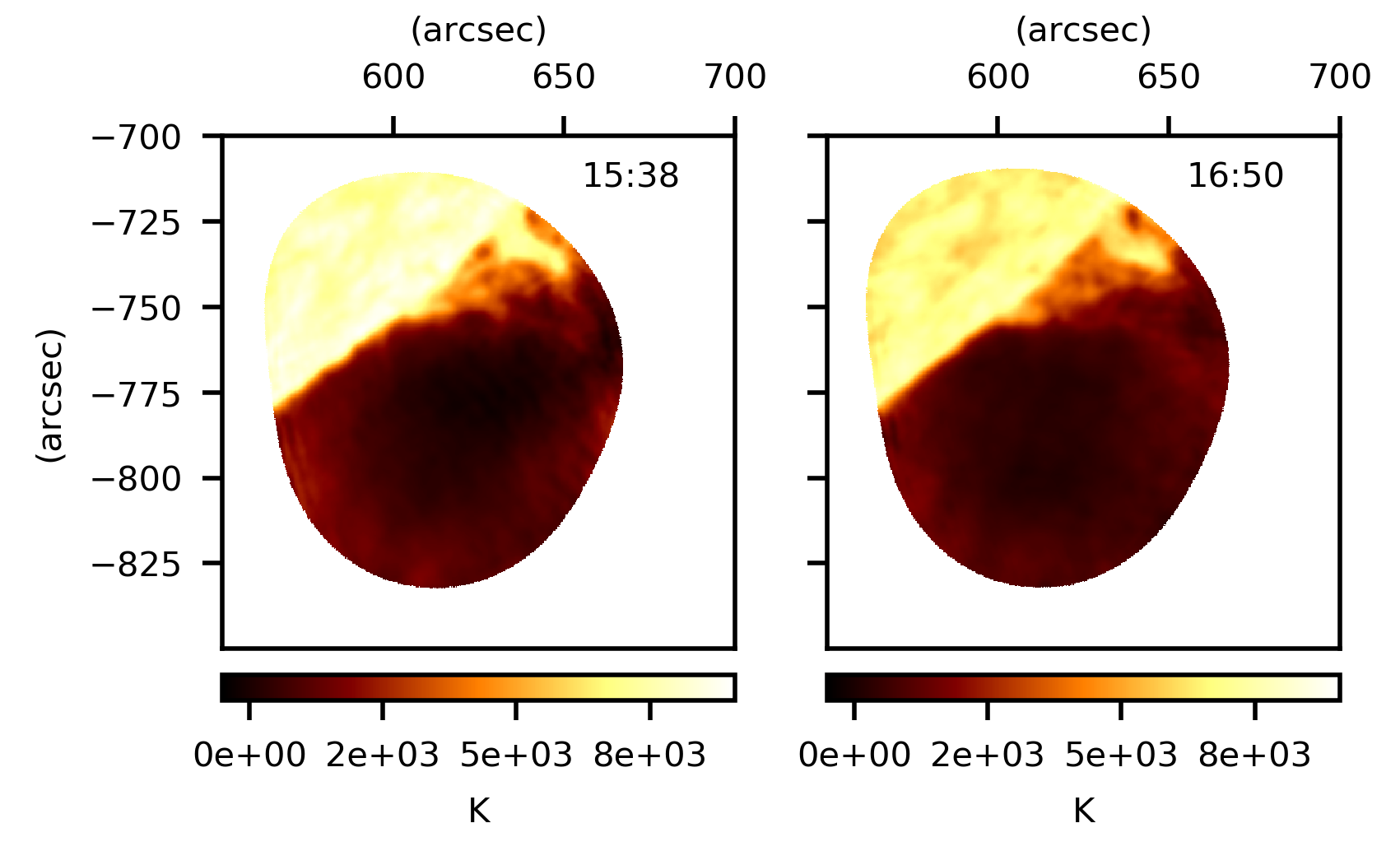}
	\caption{Absolute brightness temperature images of the prominence observed with ALMA Band 3 on  $19$ April $2018$. 
}
	\label{fig:feathered_prom}
\end{figure}

Figure~\ref{fig:feathered_prom} shows maps of absolute brightness temperature from the feathered ALMA data. 
The brightness temperatures for the prominence spine are 
between $6000$ and $7000$~K. 
Further from the limb there is some brightness temperature enhancement in the background corona.
These are likely to be caused by imaging artifacts caused by the step-function-like brightness temperature variation of the solar limb and the finite sampling of the \emph{u-v} plane.
Artifacts like these have been previously observed in other off-limb ALMA observations \citep{Shimojo2017, Yokoyama2018}.

{We coaligned the MSDP and ALMA observations closest in time to study the relation between the observed (Fig.~\ref{fig:feathered_prom}) and estimated (Fig.~\ref{fig:prom_tau_est}, right panels) brightness temperatures. Figure ~\ref{fig:tbscatter} shows a very good correlation between the two for a gas temperature of $T=8000$~K.
\begin{figure}
	\centering
	\includegraphics[width=0.9\columnwidth]{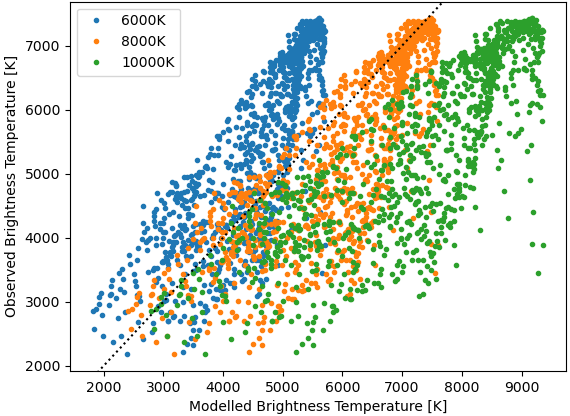}
	\caption{Observed brightness temperature from ALMA (15:38) \textit{vs} that modelled from H$\alpha$ MSDP data (15:35) in the prominence. The modelled brightness temperatures are estimated for three gas temperarures (6, 8, and 10~kK) as in Fig.~\ref{fig:prom_tau_est}. The dotted line shows when the observed and estimated brightness temperatures are equal.}
	\label{fig:tbscatter}
\end{figure}
The scatter in the plots could be explained by a non-uniform gas temperature throughout the prominence \citep[e.g.][]{2017SoPh..292..130R} and/or by a filling factor less than 1 \citep{1995SoPh..156..363I}.}

Through comparison between the Band~$3$ interferometric image (Figure~\ref{fig:interferometric_prom}) and the H$\alpha$ integrated intensity image (Figure~\ref{fig:Halpha_integrated_intensity}) it is clear that the prominence has a very similar morphology in both observations. The prominence spine is clearly visible, with what looks to be some barb-like structures at the sides, although the ALMA FOV only catches the  southward side of the full prominence observed in H$\alpha$. 
This similarity suggests  that the emission in ALMA Band~$3$ is formed either by the same cool, dense material which forms the H$\alpha$ emission; or in a closely fitting hot plasma sheath around the cool, dense core; or it is formed by the contribution of multiple, unresolved hot and cool fine-structures.

\begin{figure}
	\includegraphics[width=0.9\columnwidth]{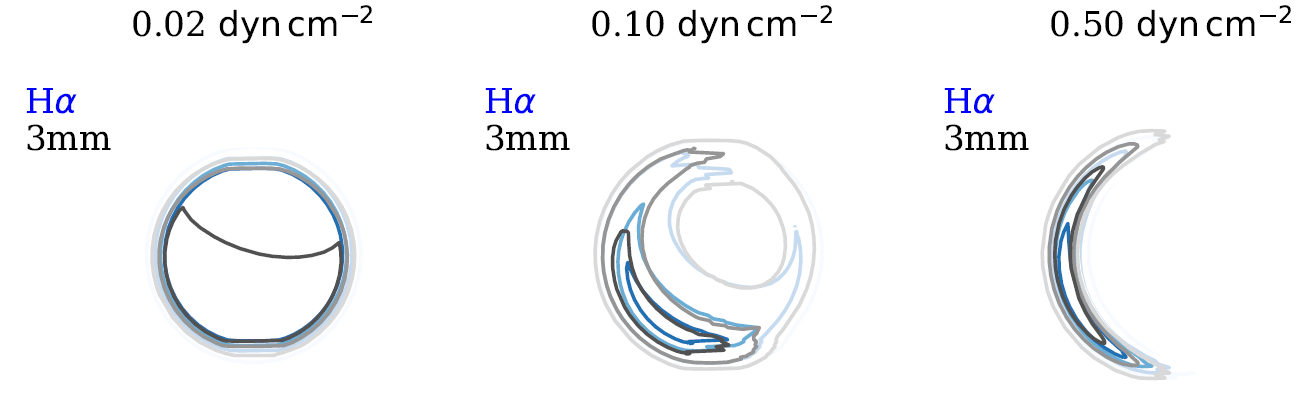}
    \caption{Contour maps for contribution functions for H$\alpha$ compared to the continuum at 3~mm, for multi-thermal prominence models at different pressures. The LOS is directed such that the observer is to the left of each cylinder. The blue contours show H$\alpha$ while the black contours describe the mm-continuum. The contour levels from light to dark correspond to {20, 40, 60,80, 100}\% of the maximum of the contribution function in the model prominence.}
    \label{fig:contrib}
\end{figure}

To address this question we computed the contribution functions at 3~mm and in H$\alpha$ using the  2D non-LTE cylindrical radiative transfer models  used in \citet{2017SoPh..292..130R}.
In figure~\ref{fig:contrib} we see a cross-section of a horizontal cylinder (radius 1000~km) located above the solar disc and observed from the left side. Optically thick emission will be seen on the left-most parts of the cylinder cross-section (right panel, higher pressure). An enhanced contribution function  at the bottom of the cross-section (left panel, lower pressure) reveals the important role played by the incident radiation coming from the solar disc underneath. 
Figure~\ref{fig:contrib} shows that the contribution functions at 3~mm and in H$\alpha$ have clear and consistent overlap across a wide range of prominence pressures. 
The larger width of the mm-continuum contribution function compared to that of H$\alpha$ suggests a contribution from hotter plasma at the base of the PCTR.

\subsection{ALMA--AIA}\label{sec:AIA_coaligned}

\begin{figure}
	\centering
	\includegraphics[width=\linewidth]{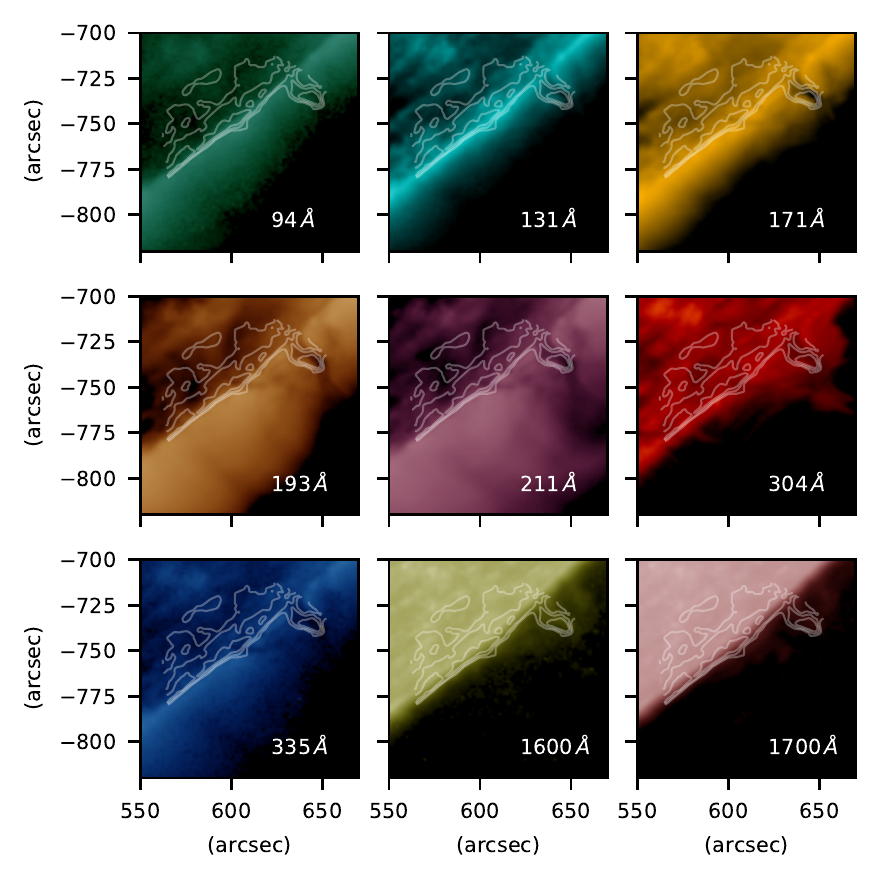} 
	\caption{Co-aligned time-averaged images of the $19$ April $2018$ prominence observed with ALMA and AIA during the second ALMA observing block. 
		The ALMA image is displayed as the white contours overlaid on the AIA images.
		The contours show the bright parts of the synthesised image with relative brightness temperatures of 1000, 2000, and 3000~K.}
	\label{fig:ALMA_AIA_coaligned}
\end{figure}

Our aim here is to compare the morphology of the prominence in the mm-continuum to the various AIA bands which are each produced at different layers within the solar atmosphere. 
For this analysis we use time-averaged AIA images for each band, as the $3$~mm images in Figure~\ref{fig:interferometric_prom} were produced across the whole time range of both observations. 
The image chosen to co-align with AIA corresponds to ALMA Block~$2$ data.
To co-align the images from the two instruments the FOV of the data sets were set to be equal. 
An example of the co-aligned images can be seen in Figure~\ref{fig:ALMA_AIA_coaligned}, where the ALMA observation is displayed as a contour map on top of the various time-averaged AIA band images. 

Figure~\ref{fig:ALMA_AIA_coaligned} shows that the contours of the Band~$3$ prominence spine follow clearly the dark structures observed in AIA  $171$, $193$ and $211$~\AA. 
This finding agrees well with previous knowledge of the optical thicknesses of ALMA Band~$3$ and in the Lyman continuum at $195$~\AA. 
The optical thickness of Band~$3$ emission is similar to that of the H$\alpha$ at line centre, and it has been previously shown by \citet{2008ApJ...686.1383H} that the opacity of the Lyman continuum at $195$~\AA\ is comparable to that of the line centre of H$\alpha$.
The bright prominence structure in $304$~\AA\ is however significantly wider than that observed in either the Band~$3$ or H$\alpha$ images. 
This is not surprising as most of the emission at $304$~\AA\ is expected to be formed under optically thick conditions in prominences, thus allowing fine-structures away from the main body of the prominence to be more easily observed. 
The $94$~\AA, $131$~\AA, $335$~\AA, $1600$~\AA\ and $1700$~\AA\ data show similar results to $171$, $193$ and $211$~\AA, however the prominence signal is significantly fainter in these channels. Emission at 1600 and 1700~\AA\ above the limb should be formed by cool lines ($10^4-10^5$~K), since the metallic continua from the photosphere cannot contribute here \citep{2019ApJ...870..114S}.
The dark structures observed in  AIA  $193$ and $211$~\AA\  follow the bright Band~$3$ prominence closer than the other AIA bands.

\section{Discussion and Conclusions}\label{sec:prom_conc}

This paper presents the first high resolution ALMA interferometric observation of a solar prominence. 
Comparison between the ALMA Band~$3$ and H$\alpha$ images reveals that both forms of emission present  similar morphologies, with both  showing a clear prominence spine and barb-like structures. 
This suggests that the mm-continuum at Band~$3$ must be formed from the same cold, dense plasma which is known to emit in H$\alpha$, or from hotter material contained within a similarly shaped structure. 
This could either be explained by a structure with a cool, dense core surrounded by a tight hot plasma sheath; or a structure consisting of multiple unresolved cold and hot plasma fine structures \citep{2021A&A...653A...5P}. 
This is consistent with the contribution function maps computed for H$\alpha$ and $3$~mm, where both types of emission have largely overlapping formation regions.

The H$\alpha$ integrated intensities are  used to estimate the optical thickness in the mm-continuum. 
We find that the maximum Band~$3$ optical thickness is $\approx 2$, which is too low for the brightness temperature to be used as a direct proxy for the electron temperature without further modelling.  
This optical thickness and the view of the prominence in Band~$3$ are in good agreement with what would be expected for a solar prominence from the models of \citet{2017SoPh..292..130R}.
Fully optically thick prominences would be expected to show brightness temperature enhancement 
through predominantly PCTR material. 
As no such brightening is observed the prominence must either be optically thin for Band~$3$ in these LOSs crossing high temperature material, and/or this material is unresolved.
{The latter is strongly supported by observations and simulations, which estimate the PCTR widths to be on the order of $0.1\arcsec - 0.3\arcsec$ \citep{2015ApJ...806...81A}.}
The brightness temperature in ALMA Band~3 (100~GHz) lies within $\approx6000-7000$~K. {Our measured and estimated brightness temperatures correlate well for a gas temperature of $T=8000$~K. The range of measured brightness temperatures} is in agreement with previous prominence modelling \citep[e.g.][]{2015SoPh..290.1981H,2017SoPh..292..130R}. 
{These results are also consistent with earlier observations by e.g. \cite{1993A&A..274L..9H,1993ApJ..418..510B} at similar wavelengths.} 

The Band~$3$ prominence observation was also compared to co-aligned images from each of the nine AIA bands. 
The AIA bands which appear to be most highly correlated with the mm-continuum emission are the $193$ and $211$~\AA\ bands, where the bright ALMA prominence spine appeared as clear dark structures against the bright surrounding corona. 
The $304$~\AA\ prominence is also clearly visible, however, the extent of this structure is considerably larger than in the corresponding mm-continuum image. 
This indicates that there is a significant amount of unresolved material which is optically thin in the mm-continuum beyond the core of the prominence spine/barbs observed in the ALMA Band~$3$ image.

Future observations of prominences with ALMA will be improved when Bands~$1$ and $2$ are available for solar observations. The prominence plasma is optically thicker at these wavelengths,  allowing more regions to be probed.
Knowledge of the temperature structure at different layers within the prominence will help improve understanding into the prominence energy balance and how they are sustained within the solar corona. 

\section*{Acknowledgements}
We thank the referee for their  comments.
Support from STFC grants ST/T000422/1 (NL, LF), ST/N504075/1 (ASR), ST/K502005/1 (PJL) and ST/S505390/1 (AWP) is gratefully acknowledged.
P.A. acknowledges STFC support from Ernest Rutherford Fellowship grant number ST/R004285/2.
PJAS acknowledges support from CNPq (contract 307612/2019-8) and FAPESP (2013/24155-3).

\section*{Data Availability}

This paper makes use of the following ALMA data: ADS/JAO.ALMA\#2017:1:01138.S. ALMA is a partnership of ESO (representing its member states), NSF (USA) and NINS (Japan), together with NRC (Canada), MOST and ASIAA (Taiwan), and KASI (Republic of Korea), in cooperation with the Republic of Chile. The Joint ALMA Observatory is operated by ESO, AUI/NRAO and NAOJ.



\bibliographystyle{mnras}
\bibliography{alma} 

\bsp	
\label{lastpage}
\end{document}